# Foundry's Perspective on Laser and SOA Module Integration with Silicon Photonics

James Y. S. Tan, Shawn Xie Wu, Yanikgonul Salih, Chao Li, and Guo-Qiang Lo

*(Invited)*

*Abstract*—Silicon photonic integrated circuit (PIC) builds on the demand for a low cost PIC approach from established silicon-based manufacturing infrastructure traditionally built for electronics. Besides its natural abundance, silicon has desirable properties such as optically low loss (at certain critical wavelengths), and small form factor to enable high density scaled-up optical on-chip circuitry. However, given that silicon is an indirect bandgap material, the platform is typically integrated with other direct bandgap (e.g., III-V semiconductor) platforms for on-chip light source. An effective solution to integrating light source onto silicon photonics platform is integral to a practical scaled-up and full-fledged integrated photonics implementation. Here, we discuss the integration solutions, and present our foundry's perspective toward realizing it.

## I. INTRODUCTION

Silicon photonics is increasingly gaining commercial traction. It is already widely used in many applications, such as telecommunications, biosensing, remote sensing (LiDAR), and computing. It initially leverages the maturity of CMOS (complementary metal-oxide-semiconductor) processing techniques in the electronics industry and is today a growingly distinct field on its own right — spanning billion-dollar market share. As the industry grows, development routines are being simplified much to the practical convenience of users/customers. These include process design kit (PDK) and packaging solutions. Open-access photonics foundries such as Advanced Micro Foundry (AMF) provides PDK device library to enable silicon photonics designers to conveniently adopt PDK devices for their respective applications without the need to develop individual devices common to the platform. However, while silicon photonics PDK provides a library comprising a plethora of silicon photonics devices, the lack of direct bandgap in crystalline silicon leads to the typical exclusion of on-chip light sources from the library. These light sources include lasers and semiconductor optical amplifiers (SOAs). Although this has been a key bottleneck that impedes full deployment of silicon photonics devices, there are currently two main general approaches to address this: laser/SOA integration onto silicon platform via packaging [1], and the more ambitious silicon-based lasers/SOAs [2]. The former is today regarded as the more practical approach and is adopted by many.

## II. INTEGRATION TECHNOLOGY

There have been multiple integration process flows for laser and SOA module integration with silicon photonic integrated circuit (PIC). As summarized in Table 1, these process flows can proceed in three major directions i.e., hetero-epitaxial integration, heterogeneous integration, and hybrid integration. These approaches have their own respective advantages and disadvantages. The choice of integration approach largely depends on the application requirement, such as integration scale, cost, and quantity of laser/SOA to be integrated onto the silicon platform.

TABLE 1. GENERAL SOLUTIONS TO LIGHT SOURCE INTEGRATION

| Integration process flow | Approach |
| --- | --- |
| Hetero-epitaxial integration | Germanium or germanium-tin laser [3] |
| | Epitaxial III-V material on silicon laser [4-7] |
| | Rare-earth-doped $Al_2O_3$ laser by external optical pumping source [8] |
| Heterogeneous integration | Molecular bonding [9] |
| | Adhesive bonding [10] |
| | Micro-transfer-printing bonding [11] |
| Hybrid integration | Passively aligned, flip-chip bonding [12, 13] |
| | Photonic wire bonding [14, 15] |

*A. Hetero-epitaxial integration*

In hetero-epitaxial integration, suitable direct-gap material as optical gain material for laser/SOA emission is grown on silicon-compatible platform. Such 'monolithic integration' of laser/SOA and silicon waveguides — built on a single silicon base material - hints at scalability and economies of scale. Of the various direct-gap materials, there is typically a tradeoff between material compatibility and lasing performance between germanium-based laser and III-V materials. While germanium-based laser (such as germanium-tin laser) offers greater integrability due to its fabrication line compatibility on silicon photonics platform, it lags behind III-V materials in terms of lasing performance (e.g., III-V material-based lasers require lower operating power and have higher side mode suppression ratio SMSR).

Of the various III-V hetero-epitaxial methods, epitaxially grown III-V quantum dot lasers is relatively promising due to the method's greater tolerance to defects and temperature drift [5, 6], which are familiar issues in epitaxially grown lasers. The increased tolerance can be attributed to the three-dimensional carrier confinement in quantum dots which discretizes (and therefore enhances) the density of states for optical emission in lasers/SOAs. As a result of such confinement, the approach promises lower lasing threshold and linewidth enhancement [7].

Although there have been notable developments in III-V hetero-epitaxial integration on silicon [16], the material remains challenging to be grown directly on silicon-based platform as III-V material inherently has larger lattice constant mismatch with silicon. The integration process involves stringent fabrication requirements, needs larger temperature budget and substantially thicker buffer layer for compensation of lattice constant mismatch and defect reduction. There is currently no hetero-epitaxial integration solution that can effectively address the tradeoff.

*B. Heterogeneous integration*

Besides combining laser/SOA material with silicon PIC material during early fabrication stage (via epitaxial growth of laser/SOA material onto silicon PIC material), laser/SOA and PIC can also be integrated during intermediate fabrication stages. In such 'heterogeneous integration', optical gain material and silicon-based optical waveguides are physically bonded together via optical coupling based on adiabatic coupling or mode overlapping method. Three major approaches to this are molecular bonding, adhesive bonding and micro-transfer printing. Molecular bonding is a direct bonding process using van der Waals forces to attach two heterogeneous materials with strong bonding strength. Although wafer-level process for molecular bonding is available through wafer-to-wafer and die-to-wafer methods for medium- and large-scale manufacturing, its process tolerance is very small and is typically limited to silicon-based chips with smaller footprint that require fewer number of lasers/SOAs due to the significantly higher cost of III-V materials. On the other hand, adhesive bonding, which uses adhesive material (polymer or metal) to bond different materials together, has larger fabrication tolerance on surface contamination and roughness, requires lower temperature for bonding process (thus involves lower possibility of introducing structural damage to III-V lasers/SOAs), and offers higher scalability with silicon wafers compared to molecular bonding. However, limitations in optical coupling efficiency is often an issue in adhesive bonding due to the following reasons: i. the thickness of adhesion polymer or metal is difficult to control, ii. adhesion polymers generally have lower thermal conductivity, and iii. it is common for metal adhesives to introduce optical loss and metal contamination. Micro-transfer-printing bonding is a distinct heterogeneous integration approach that uses a sacrificial transfer stamp (e.g., polydimethylsiloxane PDMS) to bond fabricated laser/SOA onto silicon chips. While the method enables manipulation of lasers/SOAs with high-accuracy and massively parallel/high throughput assembly of lasers/SOAs onto silicon chips, its requisite 'release-and-stamp' process onto III-V material increases process complexity and integration cost with occasional repeatability and reliability concerns. In general, because these heterogeneous integration approaches need additional back-end customized manufacturing process and tools for III-V materials such as additional III-V material etching, and cleaning capability, the approach is less compatible for current standard commercial silicon photonic foundry.

*C. Hybrid integration*

Whereas the combination of laser/SOA and silicon PIC materials in hetero-epitaxial and heterogenous integration occur during fabrication stage, the materials are combined after fabrication (i.e., during packaging stage) in hybrid integration. In the post-fabrication integration, fabricated III-V material-based lasers/SOAs are integrated onto silicon chips by using either edge coupler or grating coupler to couple light into the chips. The approach enables good die comprising laser/SOA to be selectively picked for the integration process – thus ensuring a significantly higher device performance for both laser/SOA and silicon chips, compared to other integration solutions. In AMF, the hybrid integration approach is adopted. We developed a hybrid integrated platform on 200-mm silicon-on-insulator (SOI) wafer, as shown in Figure 1.

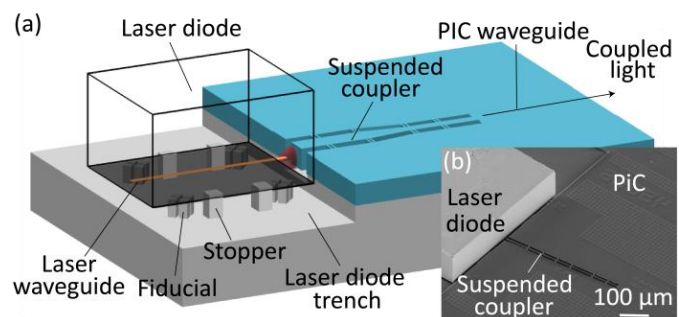

Fig. 1. (a) Schematic of example laser-PIC integration. (b) SEM image of laser diode bonded on silicon PIC.

To realize the integration, additional alignment marks are patterned for passive alignment of laser/SOA onto the silicon chip, and the stopper is defined to hold the laser/SOA and control the height of silicon chip to realize maximum optical mode overlapping. The shallow trenches ensure there is sufficient physical space to place laser/SOA onto the silicon chip. Using under-bump metallurgy (UBM) and solder bump, electrical trace and strong permanent bonding between laser/SOA and silicon photonic chips can respectively be achieved. This enables integration of III-V material such as GaAs, lnP, GaN for laser/SOA and also other materials for isolator or high-speed modulator (such as lithium niobate, $LiNbO_3$).

## III. HYBRID INTEGRATION DESIGN CONSIDERATIONS

Among the various optical coupling methods via hybrid integration, low profile lateral scheme with laser/SOA butt coupling to silicon waveguide through spot-size converter (SSC) is a promising optical coupling method. The approach, which AMF mainly pursues, offers the advantage of providing the freedom for individual optimization and fabrication of both laser/SOA and silicon photonics chips on different material platforms prior to the integration, and spatially freeing up the top surface of the silicon photonics chips for electrical and thermal connections to the active device. Fig. 2 shows the use of suspended edge couplers for hybrid integration of laser diode and PIC. Compared to conventional inverse tapered edge couplers, suspended edge couplers enable efficient optical coupling of laser/SOA beam with mode field diameter (MFD) > 3 µm although this comes at the expense of lower mechanical stability and additional fabrication process (i.e., isotropic etching).

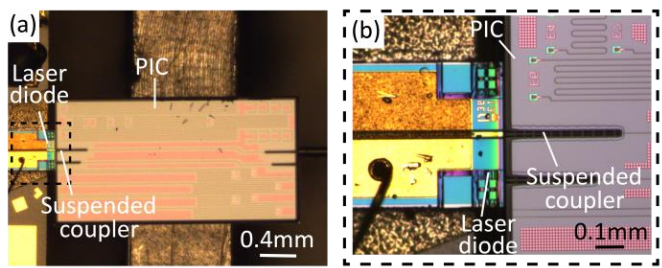

Fig. 2. (a) Optical micrograph of hybrid integration of laser diode and PIC, which uses suspended coupler for optical coupling. (b) Close-up view of coupling area. Micrograph corresponds to that in (a).

The assembly of laser/SOA module with PIC — while conceptually straightforward — entails elaborate processes and techniques to attain optimal optical coupling between the two. As these processes are often overlooked, it is important to delineate and set out the principal considerations essential to realizing laser/SOA-PIC integration. In the following, important design considerations and guidelines are described.

### A. Mechanical alignment and throughput

Mechanical alignment between the laser/SOA and PIC, in addition to individual performance of the coupling components (i.e., laser/SOA and PIC optical coupler) are essential for an effective laser/SOA-PIC integration. Mechanical alignment approaches can generally be divided into two: passive and active alignment. In passive alignment, laser/SOA and PIC are aligned manually or automatically using camera systems to determine the placement position of laser/SOA on PIC. To this end, alignment marks which can be visibly detected are typically introduced to ensure accurate positioning of laser/SOA for optimal coupling and increased throughput. This includes fiducials and Vernier scale structures for in-plane ($x$- and $y$- axis) alignment; as well as deep trenches and mechanical stops for vertical ($z$-axis) alignment. Here, fiducials and Vernier scale structures serve as points of reference to allow placement equipment to accurately position laser/SOA onto PIC, while deep trenches and mechanical stoppers are elevated structures on PIC to guarantee vertical placement of laser/SOA on PIC's deep trenches and mechanical stoppers due to high vertical alignment overlay accuracy and consistent vertical heights of the stoppers. These alignment structures ensure repeatable and reliable laser/SOA-PIC placement so that light emission centered on laser/SOA and reference plane on PIC are accurately aligned.

TABLE 2. LASER/SOA-PIC ALIGNMENT APPROACHES

| Alignment approach | Passive | Active |
| --- | --- | --- |
| Accuracy | Sufficiently high, although relatively lower. Can be affected by limited image contrast and human error. | High. Feedback ensures optimized positioning based on maximum coupled output power. |
| Repeatability | Lower. Can be affected by visual perception variability across attempts/trials | High. Fixed alignment algorithm ensures outcome consistency |
| Throughput | Generally higher with automated bonding tool. | Lower due to feedback and iterations. Depends on optimization search space and initial approximated positon. |
| Cost | Lower due to less complex alignment system | Higher. Feedback system can be complex. |

Active alignment, in contrast, involves laser/SOA placement on PIC based on their positions at which the coupled optical power from laser/SOA to PIC is maximized. This is ascertained by continuously monitoring the coupled optical power via real-time closed loop feedback system. Although a series of predetermined motion sequences can be adopted to determine the optimal position, the use of optimization algorithms to maximize the coupled optical power is commonplace. This usually capitalizes on sub-µm piezo stage movements to precisely and automatically track laser/SOA-PIC positions with respect to their resulting coupled optical power. The right choice of optimization algorithm in an automated alignment settings can ensure quick and increased alignment throughput. Table 2 compares passive and active alignment approaches. While active alignment ensures highly accurate and repeatable laser/SOA-PIC placement, the approach comes with a generally higher cost and lower throughput due to the feedback system. However, with 1 dB alignment tolerance at ~1.5 µm in many laser/SOA-PIC optical coupling schemes along with current align and attach tools capable of invariably achieving <0.5 µm placement accuracy and <1.5 µm three-sigma accuracy (e.g., MRSI-S-HVM, SETNA FC300 and ASMPT CoS Die Bonder), passive alignment for laser/SOA-PIC bonding is already a feasible solution for mass production today.

To maintain optical alignment and assembly integrity, submounts are used. They can be formed via epoxy dispense/daubing, and soldering processes (e.g., in the form of flip-chip bonding [13, 17], solder reflow [18, 19], and eutectic die attach [19, 20]). Although epoxy offers the lowest cost, the material may cause mechanical stability issues due to outgassing, thermal expansion, loss of adhesion, shrinkage upon cure, and humidity. Lack of mechanical stability can inadvertently lead to alignment shifts that can affect optical coupling between laser/SOA and PIC. As epoxies can serve as an index matching medium to improve laser/SOA-PIC optical coupling, epoxies incorporated with particle fillers have been used to adjust the refractive index of the epoxy, and CTE to respectively improve both the optical transmittance and mechanical stability [21]. Soldering processes, on the other hand, offer better mechanical stability albeit at a higher processing cost. Nevertheless, the approach is typically accompanied with metallization failure concerns, generally due to interface reactions. While flux may be used to facilitate metal amalgamation, it may cause residue and induce oxide formation, which introduces contamination. Following soldering process, the removal of flux have been proven to extend solder lifetime.

*B. Optical coupling efficiency and tolerance*

Although the main objective of laser/SOA-PIC integration is to couple light from the laser/SOA into PIC as efficiently as possible, the optimal coupling approach ultimately rests on the specific application requirements, given the different constraints in different application settings. The requirements can include coupling efficiency, wavelength coupling bandwidth, alignment tolerance and manufacturability. Table 3 summarizes the different coupling approaches. Similar to optical fiber-PIC coupling approaches [22-30], they can be divided into lateral and vertical coupling. Lateral coupling can be achieved via waveguide extension [14, 15], spot size conversion [31-33] and adiabatic coupling [34], while vertical coupling via grating coupler [35], overlay structure [36], prism/lens, grating coupler combination [37], and mirror-reflected edge coupler

Conventional lateral coupling approach using spot-size conversion has high coupling efficiency with relatively simple structural design and high component manufacturability, but comes with low alignment tolerance of more than 0.5 dB/µm misalignment [31-33]. In particular, suspended coupler [33] — an example of the approach, enables high coupling efficiency. Fig. 3 describes the development of laser-PIC integration at AMF using tilted suspended coupler. Experimental measurements indicate 2.7 dB coupling loss at 60mA bias current. Compared to spot-size conversion [31-33], adiabatic coupling-based lateral coupling methods can achieve very high coupling efficiency of less than 0.5 dB loss, but is design parameter sensitive [34], while waveguide-extension based lateral coupling methods can achieve about ~0.4 to 2 dB loss with very high wavelength coupling bandwidth of more than 100nm but has low alignment tolerance of more than 0.5 dB/µm misalignment, and low manufacturability [14, 15].

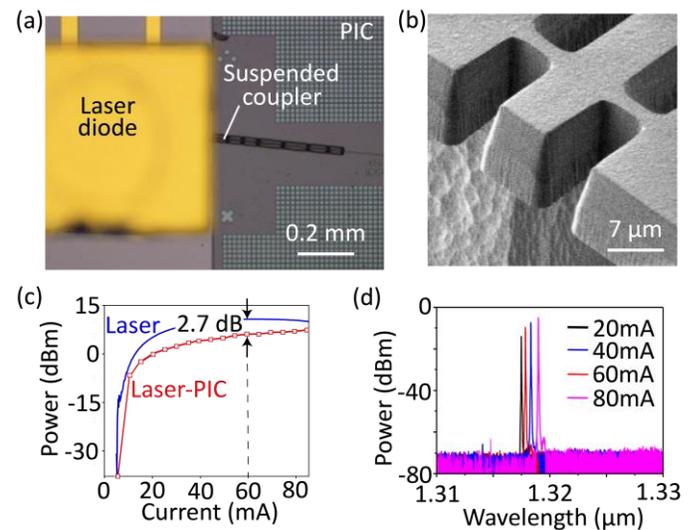

Fig. 3. (a) Optical micrograph of laser-PIC integration, with tilted suspended coupler to enhance optical coupling. (b) SEM image of coupler edge. (c) Measured optical power-current plot of laser diode (blue) and coupled laser-PIC structure via suspended coupler indicate 2.7 dB coupling loss at 60 mA bias current. (d) Measured optical spectra of bonded sample at bias current from 20mA to 80mA.

Although conventional grating coupler-based vertical coupling approaches have higher alignment tolerance, they typically come with lower coupling efficiency and coupling bandwidth [35]. Other vertical coupling approach may offer complementary features such as compact lateral footprint using overlay structures, but requires additional fabrication steps and is design-parameter sensitive [36]. There have been ongoing active development efforts to develop optical packaging-compatible platform using a combination of prism or lens and grating couplers [37]. This bodes well for further electrical signal and optical signal interfacing, but needs additional packaging steps. Other similar approach aims to increase the overall manufacturability using conventional semiconductor device fabrication via a combination of mirror-reflected edge coupler and grating coupler instead of assembling discrete optical prism or lenses [38]. This comes with additional fabrication steps.

At AMF, to increase optical power handling and mechanical stability of PIC coupling structures, tri-layer edge couplers have been developed. Fig. 4a depicts the tri-layer structure. As spot-size conversion structures typically have narrow dimensions (e.g., ~0.1µm tip width in both inverse and suspended couplers), optical field concentrated at the tip may heat up the area and induce structural failure. Tri-layer edge couplers circumvent this by distributing the optical field to a larger area (~2.5 × lattice constant), with MFD of 10.4µm in our finite-difference time-domain (FDTD) numerical simulations. As shown in Fig. 4c, measured optical spectra indicate ~2dB coupling loss at ~1.55µm wavelength. Compared to the typically free-standing suspended edge couplers, the overcladding-confined tri-layer structure is mechanically robust to shock and vibration.

TABLE 3: LASER/SOA-PIC OPTICAL COUPLING SCHEMES

| | Waveguide extension [14,15] | Spot size conversion [31-33] | | | Adiabatic coupling [34] |
|---|---|---|---|---|---|
| | | Conventional [31] | Trident structure [32] | Suspended [33] | |
| Lateral Coupling | • High coupling efficiency (~0.4 to 2 dB loss)<br>• High wavelength coupling bandwidth (>100 nm)<br>• Low alignment tolerance (> 0.5 dB/μm)<br>• Low manufacturability | • Simple structural design<br>• High component manufacturability<br>• Low alignment tolerance (> 0.5 dB/μm) | | • High coupling efficiency (~1 to 2 dB loss)<br>• Low mechanical stability and strength | • Very high coupling efficiency (<0.5 dB loss)<br>• Design parameter sensitive |
| | Grating coupler [35] | Overlay structure [36] | Prism/lens, grating coupler combination [37] | Mirror-reflected edge coupler, grating coupler combination [38] | |
| Vertical Coupling | • High alignment tolerance (< 0.5 dB/μm)<br>• Lower coupling efficiency (~3 dB loss)<br>• Lower wavelength coupling bandwidth (<100 nm) | • Compact lateral footprint<br>• Requires additional fabrication steps<br>• Design parameter sensitive (e.g., coupling length) | • Photonic packaging-integrable platform<br>• Requires additional packaging steps | • High overall manufacturability (both LD and PIC) using conventional semiconductor device fabrication methods<br>• Requires additional fabrication steps | |

As a general rule-of-thumb, laser/SOA-PIC coupling efficiency can be improved by reducing the emitted beam divergence on the laser/SOA end (e.g., via SSC), and enhancing mode field matching on the optical coupling element on the PIC end. Given the constraints in device and/or assembly settings, it is desirable to have certain tolerances to optical coupling. For example, alignment insensitive optical couplers to circumvent precision/accuracy-lacking placement settings.

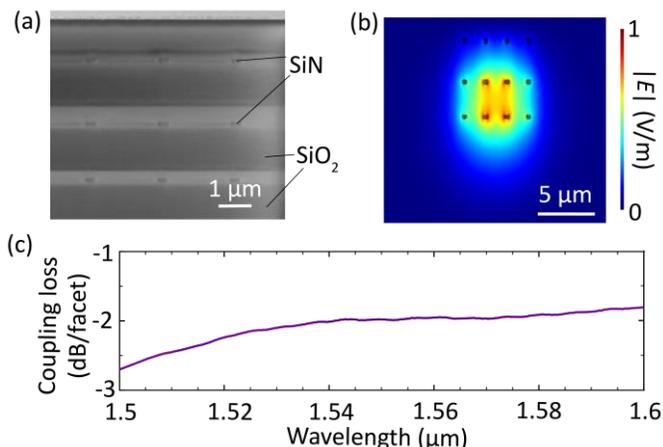

Fig. 4. Tri-layer edge coupler for mechanical fracture-tolerance and optical power handling. (a) SEM image of cross-section of SiN tri-layers. Layers are integrated to bottom Si layer (not shown) via Si inverse taper. (b) Simulated mode profile of coupler facet. (c) Measured spectra of coupler.

*C. Laser/SOA, and PIC device performance*

Laser/SOA-PIC optical coupling efficiency is highly contingent on the device performance of each individual components in the integration i.e., laser/SOA and PIC, in addition to the mechanical alignment of the assembly. Solder bonding reliability issues and laser/SOA emission wavelength shift occurrence (which affects device performance) are typically exacerbated by poor on-chip heat management. Over time, due to aging effects expedited by heat and moisture — among other compounding factors, interface reactions can dominate to result in solder bonding failure. To improve heat management, heat dissipation and isolation structures can be engineered using an interplay of locally removed buried oxide (BOX), thermal shunt with highly thermally conductive material, heat spreader connected to a thermoelectric cooler element, forced air convection heat sinks [39].

Other approaches to enhance optical coupling efficiency between the components is to enhance the coupling itself. This includes both the deposition of anti-reflection coating on laser/SOA facet to enhance optical transmission through the laser/SOA, and the use of index matching oil and epoxies to match the refractive indices of laser/SOA and PIC coupling structure to effectively reduce Fresnel reflections at both ends of the components.

*D. Aging effect minimization*

In addition to the aforementioned factors and considerations, it is equally important that the integrated structure can withstand aging effects. To weather such effects, device passivation and sealants have been used. The need for device passivation arises due to the fact that laser/SOA material such as III-V semiconductor materials (e.g., InP, InAs, and InGaAs) have high surface state density (>$10^{13}$ cm$^{-2}$) and lacks good intrinsic oxide passivation layer [40]. This reduces laser emission efficiency due to non-radiative recombination of carriers, and introduces oxidation induced states which are responsible for catastrophic optical damage (COD) and therefore reduced device lifetime. Before the deposition of anti-reflectivity coating, laser/SOA facet is coated with passivation layer to prevent oxidation and corrosion of the laser/SOA material. This renders the semiconductor surface inert so that it does not oxidize. Example of passivation layer material includes ZnSe, SiN [41].

Contamination from environment on laser/SOA-PIC structure can detrimentally affect the proper functioning and lifespan of laser/SOA-PIC coupled structure. For example, moisture from condensation can lead of electrical contact corrosion, thermal shorting due to water bridging, and optical coupling distortion in the laser/SOA-PIC coupled structure e.g, both the laser/SOA emission efficiency and laser/SOA-PIC coupling efficiency. Contamination may come in the form of prolonged moisture, pressure, temperature, and chemical reactions. To secure against moisture and chemical contamination, the coupled structure can be packaged in a protective seal. The implementation of such seals have been especially considered in cases when the PIC structure comprises sensitive microelectromechanical systems (MEMS), optomechanical, and cryogenic photonic devices. Protective seals can generally be divided into two: hermetic and non-hermetic seals.

In hermetic seals, lids are used to ensure a consistently airtight housing of the coupled structure. The hermetic lid material are typically glass, ceramic, or metal (e.g., high-$\kappa$ dielectric, eutectic metal, ultrasonically-welded metal [13, 42]). The seals eradicate contamination risks with its impervious airtight casing. In the sealing process, the laser/SOA-PIC structure is first subjected to vacuum bake-out to remove moisture or other residues, then filled with inert buffer gas (e.g., nitrogen $N_2$ or argon Ar), and finally hermetically sealed (e.g., using roll seam welded metal lids). The buffer gas in the sealed package fixes the device overcladding surrounding in the package. Non-hermetic seals, on the other hand, are less airtight and serves to limit and restrict the passage of moisture and contaminants to a considerable extent. These seals can be in the form of coatings or gap-free edge seals. Examples include silicone/rubber gaskets, epoxy and acrylic resins. Although the choice of sealing approach is principally dependent on environment settings, laser/SOA-PIC coupled structures without hermetic packaging have been widely used in data center deployments with little to no reliability issues [43]. With an increasingly economical option using hermetic sealing, hermetic seals may find more extensive use in laser/SOA-PIC coupled structures for consistent, reliable and proper functioning of the structures over an extended lifespan.

*Basic guidelines for laser/SOA-PIC integration*

These methods to achieve optimal optical coupling efficiency and tolerance, high mechanical alignment and throughput, good individual performance, and reduced aging effects varies depending on constraints such as process complexity, yield, cost, and reliability factors. As such, different combinations of designs that simultaneously address the considerations earlier entailed in this section — which are customizable — may be attempted accordingly. At AMF, we follow a set of basic guidelines for laser/SOA-PIC integration. These guidelines, summarized in Table 4, contain recommendations that are targeted at requisites to optical coupling between the components i.e., optical coupling efficiency, and mechanical alignment of the integration components.

TABLE 4. AMF GUIDELINES FOR LASER/SOA-PIC INTEGRATION

| Parameter | Guidelines |
|---|---|
| Optical coupling efficiency | i. Laser/SOA:<br>- Design SSC to reduce emission beam divergence (e.g., ~4µm MFD)<br>ii. PIC:<br>- Match mode field of PIC coupler with that from laser/SOA. Design angled edge coupler to reduce optical field back reflection. (e.g., 7° to 12°). |
| Mechanical alignment | i. Fiducials/alignment marks<br>- Include alignment marks on both laser/SOA and PIC to improve alignment accuracy<br>- Optional laser/SOA etch facet for additional alignment.<br>ii. Stopper pads<br>- Desirable stopper size to provide stable structural support<br>- Design laser/SOA and PIC stack-up structures for vertical alignment accuracy.<br>iii. Shallow trench<br>- Shallow trenches on PIC wafer for laser/SOA placement |

## IV. HYBRID INTEGRATION PROCESS CONSIDERATIONS

*A. Metallization*

As lasers/SOAs are light sources that operate through stimulated emission, additional energy has to be pumped to the gain medium of these light sources to produce population inversion. Electrical pumping is commonly used for the pumping process, in which population inversion can be induced when electrical current sufficiently passes through the gain medium to result in a net optical amplification inherent to these light sources. Fig. 5 briefly describes the metallization aspect of laser/SOA-PIC coupling via flip-chip process. On the PIC, the wirebonding pad enables external electrical current to be supplied to laser/SOA via bond pads and metallization layers. Flip-chip bonding enables connection between electrical pads on both laser/SOA chip and PIC chip.

Supplying electrical current to lasers/SOAs require additional metallization considerations as metals on electrical wires do not typically bond directly with the semiconductors in the gain medium of diode-based lasers/SOAs. Besides providing the required electrical contact, the bond needs to have low thermal and electrical contact resistance and be able to withstand high mechanical force (e.g., shear, stress, vibration and shock). The metallization approach to enable bonds with such features typically comprises metallization layers. Table 5 provides a summary of the layers.

TABLE 5: TYPICAL METALLIZATION LAYERS

| Layer | Purpose | Typical materials |
|---|---|---|
| Adhesion | Bonds semiconductor with subsequent metal layers | Ti |
| Barrier diffusion | Prevents diffusion of adhesion layer to subsequent metal layers, and of subsequent metal layers to semiconductor | Ni [44] Pt [45-48] |
| Capping | Prevents oxidation of adhesion and barrier diffusion layer, complete intermetallic formation during reflow, and enables wetting of prior layers with solder layer | Au Pt [49] |
| Solder | Conjoin metal alloy on semiconductor with metal contact | AuSn [42, 49-51] SnAg [52] |

While adhesion layer is essential to provide adhesive bonding via molecular contact between the layer material with both the underlying semiconductor and electrical wire, it is still susceptible to natural chemical reactions arising from exposure to surrounding material and ambient atmosphere e.g., diffusion and oxidation. As a result, additional metal layers that prevent these processes while maintaining electrical connection has to be deposited as well. Adjoining the adhesion layer is the barrier diffusion layer. The barrier diffusion layer is inherently inert and non-reactive to the adjacent material to prevent diffusion between adhesion layer, other metallization layers, and semiconductor. This ensures the constituents from these different materials do not undesirably contaminate, dissolve and/or induce substantial intermetallic formation in the layers. Nickel (Ni) and platinum (Pt), which have low diffusion coefficients, are typically used as materials of choice for barrier diffusion layer [44-48]. In particular, Pt is commonly used as a diffusion barrier layer to prevent penetration of Au into Ti adhesion layer (which has good adhesion with both Si and III-V semiconductors). Other layer to the stack is the capping layer, which is deposited to prevent oxidation of both adhesion and barrier diffusion layer. Noble metals such as Au and Pt are materials of choice for capping layer as they have outstanding resistance to oxidation [49]. Being extremely inert, they do not deteriorate or form alloys or corrosion products with other materials even at high temperatures. Overall, these adhesion/barrier diffusion/capping layers may be deposited in varying sequential orders.

Finally, to ensure a relatively fastened and intact mechanical bond between the electrical wire and the metal layers adhered to silicon on PIC, the solder layer is deposited. The mechanical bond in the form of solder joint is realized by depositing solder material on the capping layer before conjoining the metal contact to establish electrical current supply to laser/SOA diode.

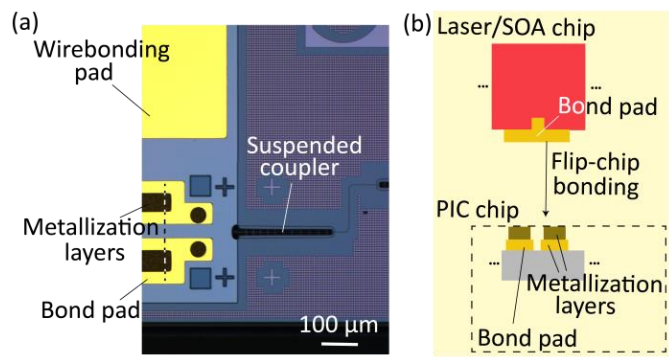

Fig. 5. (a) Micrograph of PIC for laser/SOA-PIC integration. Metallization layers and bond pad enables electrical current to be sent to laser/SOA. Wirebonding pads are electrically routed (not shown) to bond pads. (b) Schematic of flip-chip bonding process for laser/SOA-PIC integration.

*B. Solder attach methods*

The deposition of adhesion, barrier diffusion, and capping layer on lasers/SOAs is commonly realized using conventional semiconductor device fabrication processes [13], given the typically nm-thick deposition that these layers require. The solder layer, on the other hand, may be substantially thicker to accommodate conjoining electrical wires which may be µm-thick to the adhesion/barrier diffusion/capping layers on lasers/SOAs. Restricting to conventional semiconductor device fabrication processes for solder layer deposition may impede volume production, due to the low deposition rate of the processes. The tradeoffs of the different methods to attach solder layer onto adhesion/barrier diffusion/capping layers is summarized in Table 6.

TABLE 6: SOLDER ATTACH METHODS

| Method | Key advantages | Disadvantages |
|---|---|---|
| Preform [53-55] | Straightforward process | Poor composition control, low volume |
| Paste [56] | High volume | Poor composition control |
| Sputtering [57]/ evaporation [17, 58] | Excellent composition control | Low volume, expensive, process sensitive |
| Electrochemical plating (ECP) [13, 49, 59] | High volume, good composition control | Process sensitive |

In general, these methods can be categorized to pre-configured approach and direct film deposition approach. The pre-configured approach involves preparing the solder material in a handy form beforehand such that the solder attach process can be readily carried out typically by heating the solder material, e.g., preform and paste methods [44, 53, 56]. Prior to the attach process, the solders are respectively in solid structure of solder shaped in a predetermined manufactured form and powdered solder dispersed in a viscous flux paste in preform and paste methods [44, 53, 56]. Direct film deposition approaches, on the other hand, entails depositing films of solder material in a tightly controlled environment, e.g., sputtering/evaporation and electrochemical plating (ECP) [17, 49, 56-59]. In sputtering/evaporation method, chamber with controlled parameters is used via self-sustaining plasma or electron beam for solder deposition. In ECP, electrochemical reduction of cations of solder elements using direct electric current is used to coat solder film onto the target metal contacts [49]. Direct film deposition approaches are widely adopted for metallization in lasers/SOAs due to their relatively better composition control compared to pre-configured approach. Of the common direct film deposition approaches, sputtering/evaporation and electroplating are respectively opted when excellent composition control and high-volume manufacturability are required.

TABLE 7: SOLDER METALLIZATION METHODS ADOPTED BY AMF

| Solder metallization option | Au/AuSn | Cu/SnAg |
|---|---|---|
| Deposition process | Evaporation | ECP |
| Solder melting point (°C) | >280 | ~220 |
| Volume production | Small | Potentially high |
| Cost | Higher (due to Au) | Relatively low |
| Features | Different bonding temperatures can be adopted to eliminate possible cross effects between bonding processes of laser/SOA chips and electronic chips (e.g.,transimpedance amplifiers (TIAs), drivers). | Less prone to device damage due to lower bonding temperature (e.g., ~220°C for Cu/SnAg) and bonding time. However, CTE between the bonded materials may introduce unwanted mechanical strain. |

At AMF, the solder material AuSn and SnAg, with bonding temperatures of respectively >280°C [42, 50, 51] and ~220°C are used. Table 7 shows a comparison between the two options. While AuSn solder is deposited via evaporation, SnAg solder is formed via ECP. As deposition rate by evaporation is low (at most ~several μm/min), solder formation via ECP — with deposition rate ~tens of μm/min [59] — is more suited for volume production. Another factor to be considered for the selection of solder material is electronic chip integration. Besides the light source bonding, the electronic chips, e.g., TIA, drivers, would also need to be attached on the silicon photonics chips. Having substantially different bonding temperature could eliminate the potential cross effects between bonding processes of light source chips and electronic chips.

## V. Reliability

Laser/SOA-PIC structures comprise multiple structural parts that collectively operate together to optically couple coherent light to PIC. The device quality relies on multiple compounding factors. They commonly include device material, assembly/packaging parts, and the less apparent interface reactions. A good understanding of failure mechanisms of the device would help ensure device integrity through manufacturing processes and operating conditions.

### A. Laser/SOA-PIC performance parameters

Device failure can be accessed by first establishing then measuring the performance parameters of the device. There are several methods to access laser/SOA and/or laser/SOA-PIC device performance. They are:

i. LIV characteristics.

The operating point of a laser/SOA is typically determined from the LIV curve, where L, I, and V are respectively optical output power, monitor photodiode current, and voltage. Examples of tests from which the LIV curve is commonly plotted to ascertain the direct current characteristics of the laser/SOA include verifying the forward voltage drop across the laser/SOA as current is swept (i.e., forward voltage test), measuring the optical power as drive current is increased (i.e., light-intensity measurement), determining the drive current at a rated optical power and threshold current at which lasing begins (i.e., lasing threshold current test), determining the current generated in the back facet of the laser/SOA (i.e., back-facet monitor diode test), verifying the proportionality of relationship between the laser/SOA drive current and optical power (i.e., kink test/slope efficiency measurement), and determining the range of temperatures at which the laser/SOA operates optimally (i.e., temperature test).

ii. Spectral profile.

The spectral profile of laser/SOA provides the range of wavelengths at which the laser/SOA operates given a particular device settings. Examples of relevant tests include verifying bandwidth/wavelength tuning range of laser/SOA and checking the wavelength tunability at the specified tuning range (e.g., multi-hop-free tuning in single-frequency lasers).

iii. Spatial profile

Laser/SOA beam spatial profile provides information about the irradiance profile of the laser/SOA beam that exit the laser cavity. The profile is crucial to ensuring efficient laser/SOA-PIC coupling as deviations from characterized profile may result in beam mode mismatch, which decreases optical coupling efficiency between laser/SOA and PIC. Examples of relevant tests include verifying the laser/SOA emission pattern at varying drive current and other conditions, and retrieving the beam divergence angle of the laser/SOA. Inconsistent emission pattern may indicate laser/SOA facet structural inconsistency or damage.

iv. Pulse characterization (for pulsed lasers)

For pulsed lasers, measurement of output optical power as a function of time (temporal pulse shape) provides information about time-dependent laser/SOA emission and emission spectrum.

Measured deviations from the typical test values reflect possible anomalies in the device, which can indicate device failure. These failures can generally be classified as infant mortality failure, external factor failure, and wearout failure. Infant mortality failure is caused by defects introduced during manufacturing process or intrinsic semiconductor defects, for example during epitaxial growth of semiconductor. The failure rate typically reduces significantly with device usage. External factor failure is caused by external environmental factors e.g., due to occurrences of current surges or electrostatic discharge (ESD) occurrences. The failure rate generally remains constant with device usage. In contrast, wearout failure is caused by growth of non-radiative/optically absorbing defects in the active region of laser/SOA typically after long operating hours. Examples of wear out failure is the general degradation of laser/SOA material and facet coating. This failure increases significantly with device usage.

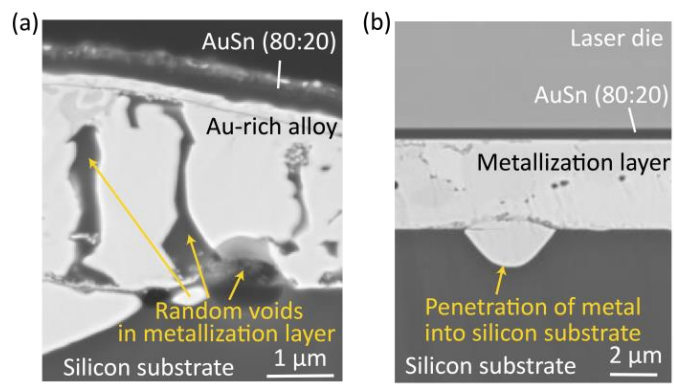

Fig. 6. Reliability issues in metallization layer. (a) Random voids. (b) Metal diffusion into silicon substrate.

The result of such failure may manifest in a steadily decreasing efficiency or instant complete failure. Failures that result in a steadily decreasing efficiency include those due to dislocation growth in semiconductor lattice, facet oxidation, electrode, electrical link and heat sink degradation. A typical example of instant complete failure is catastrophic optical damage (COD) which occurs when the laser is overloaded with exceedingly high power density causing the junction to excessively absorb the generated optical energy to result in recrystallization of semiconductor material at the laser/SOA facet. Fig. 6 depicts the typical reliability issues in metallization layer i.e., the formation of random voids and the structural penetration of metal into silicon substrate as a result of metal diffusion.

*B. Reliability tests*

Given the possibility and risk of device failure, it is crucial to anticipate unfamiliar operational and environmental conditions, identify inherent performance and failure issues, and incorporate statistical process control (SPC) methodologies in device fabrication and perform reliability tests. While SPC methodologies ensures that the out-of-specification probability or performance inconsistency/variability of the device is kept within a certain established tolerance specification, it does not address performance inconsistency over time, and under-additional-environmental stresses. Reliability tests, in contrast, are tests specifically designed to address this. In general, there are two main approaches to testing reliability i.e., computational modelling and accelerated aging. Due to the generally multifaceted factors associated with device breakdown (e.g., pressure, time, humidity, temperature), computers have been used to simulate and study such complex systems in order to retrieve predictions regarding the stress and/or temporal extents at which the device fails. Another reliability test commonly adopted in industry (including AMF) is accelerated aging. By subjecting the device to aggravated stress conditions (e.g., of heat, humidity, vibration), significantly longer-term effects of expected levels of stress can be predicted within a shorter time period using Arrhenius equation. The equation describes the temperature dependence of reaction rates, and can be written as

$$t = Ae^{-\frac{E_a}{kT}} \qquad (1)$$

where $t$ is time, $A$ is the test constant, $E_a$ is the activation energy dependent on failure mechanism, $T$ is temperature. By taking the ratio of the reaction rates at two different temperatures (i.e., use temperature $T_1$ and test temperature $T_2$), we obtain the acceleration factor $AF$:

$$AF = e^{-\frac{E_a}{k}\left(\frac{1}{T_1} - \frac{1}{T_2}\right)} \qquad (2)$$

For example, for a device calculated to have $AF = 29$ tested at $T_2 = 100°C$ to estimate the case at $T_1 = 50°C$ for a given $E_a$: when the device is tested for 1000 hours, the estimate is valid for $29 \times 1000 = 29000$ hours. In practice, there is a limit to the test temperature $T_2$ as this may lead to an unrealistic failure if set to a value higher than that the device can withstand.

The demonstration of device survival of a series of reliability tests that conforms to certain recognized international standards is known as 'qualification'. In photonics/electronics industry, these recognized standards include those from Telcordia, Joint Electronic Device Engineering Council (JEDEC), Automotive Electronics Council (AEC), and Military Standard (MIL-STD). They can generally be divided into four grades i.e., consumer, industry, automotive, and military — in order of increasing requirements. For example, devices built for military applications require grater tolerances to harsh environments, and thus require more stringent quality control. In general, AEC Grade 1 which covers from -40°C to +125°C is sufficiently extensive and sufficient for most target applications. Although there is currently no known reliability standards specific to laser/SOA-PIC coupled structures, reliability tests for those structures have been implemented similar to those for lasers/SOAs [60]. These include bias currents [43, 61, 62], optical output power-current characteristics [61-66], capacitance-voltage characteristics [67] and lasing spectra [61, 65] under specified stress conditions (e.g., constant current and current step). Reliability tests for the coupled structure have to be considered in a holistic manner, as the integration procedure may introduce defects e,g, thermally-induced bonding defects [66, 68]. Table 8 shows the recommended test conditions for laser/SOA-PIC coupled structure that covers AEC Grade 1.

TABLE 8: RECOMMENDED TEST CONDITIONS FOR LASER/SOA-PIC COUPLED STRUCTURE

| Qualification item | Stress test | Suggested test conditions | Reference |
|---|---|---|---|
| Electrical link | Stress migration (SM) | 175, 200, 225, 250 & 275 °C; T0, 48, 100, 250, 500, 1000 hours | JEP001, JEP139 AEC-Q100 |
| | Electro migration (EM) | Constant current stressing at 150 °C – 250 °C | JEP001 AEC-Q100 |
| | Electrostatic discharge (ESD) | Refer testing guidelines | AEC-Q100-002-Rev-E-2013 |
| Mechanical integration | Wire-bond pull | Suggestion: Pre and post temperature cycling (TMCL) and unbiased damped heat (uBDH) | AEC-Q102 - Rev A |
| | Wire-bond ball shear | | |
| | Die shear test | -40ºC to 125ºC, minimum soak & dwell time 15 min., up to 1000 cycles 85ºC, 85% RH; T0, 168, 500, 1000 hours | |
| Packaging solution | Temperature cycling (TMCL) | -40ºC to 125ºC, minimum soak & dwell time 15 min., up to 1000 cycles | AEC-Q102 - Rev A |
| | Unbiased damp heat (uBDH) | 85ºC, 85% RH; T0, 168, 500, 1000 hours | |
| | High temperature storage (HTS) | Ambient temperature = 125ºC; T0, 168, 500, 1000 hours | GR-468 |
| | Low temperature storage (LTS) | Ambient temperature = -40ºC; T0, 168, 500 hours | |

Recommended reliability specifications are as detailed in the table references.

The recommended test conditions have been classified into three qualification items i.e., electrical link, mechanical integration, and packaging solution. Three of the more common stress tests for electrical interconnects are stress migration, electromigration and electrostatic discharge. Stress migration and electromigration are caused by mechanical stress gradient and diffusion of ions in the electrical link respectively; while electrostatic discharge is caused by electrostatic charge induced by high electrostatic field which may be large enough to induce damage to sensitive electrical links in a device. There are multiple ways to test the mechanical integration of the components in laser/SOA-PIC coupled structure i.e., creep (of constant mechanical strength), deformation due to monotonically increasing mechanical strength, and fatigue (of cyclic mechanical strength). The most immediate and pressing mechanical reliability tests are that of creep. They are wire-bond pull, wire-bond-ball shear and die shear test [20]. One common cause of bond failure is mismatches in coefficient of thermal expansion (CTE) [13]. As a result of different thermal expansion rate, the resulting stress can lead to fractures and cracks in the bond medium. The problem can also be attributed to interface reactions from the multiple metallization solder layers (i.e., adhesion, barrier diffusion, capping, and solder layer). Table 9 summarizes the cause of common interface reactions. Intermetallic growth, Kirkendall void formation and 'return of Au' phenomenon are the result of diffusion of atoms, ions or molecules [69-71], while black pad formation is due to deposition process reaction resulting in corrosion of critical component in the metallization layer [72]. To rigorously evaluate of quality of the bond connected to PIC (and therefore laser/SOA) for electrical supply, wire-bond pull and wire-bond-ball shear tests can be performed. Die-substrate adhesion, on the other hand, can be tested via die shear test [20].

The packaging/assembly of laser/SOA-PIC structure includes underfill, epoxy [18], solder joints adapted to enable mounting of the different components on a single platform. These assembly components are requisites to an optically aligned coupled structure, which can be undesirably disintegrated due to a myriad of reasons such as temperature cycling, humidity, elevated and reduced temperatures [13]. Example tests include temperature cycling test, unbiased damped heat for humidity, high and low temperature storage for elevated and highly reduced temperatures. Particularly on laser/SOA testing, three types of aging studies can be carried out to predict laser lifetime. They are constant current aging, constant power aging, and periodic sample testing. In constant current aging and in constant power aging respectively, electrical current and optical power (while continuously adjusting electrical current) are held constant throughout the test duration. Periodic sample testing pertains periodically reducing the temperature of a laser to test lasing action at a relatively elevated temperature. To effectively weed our devices that have likely short lives, screening methods are used by running the devices at an elevated temperature continuously over many hours (i.e., burn-in process), and identifying the change in at least one performance parameter before and after the process.

TABLE 9: CAUSE OF COMMON INTERFACE REACTIONS

| Interface reactions | Causes | Effect |
|---|---|---|
| Intermetallic growth [71] | Element migration causing the formation of intermetallic compound (with possibly weaker mechanical strength) to be formed | Solder fracture |
| Kirkendall void formation [19, 69] | Void formation due to motion of metal interface as a result of different diffusion rates of the metal atoms | |
| 'Return of Au' phenomenon [70] | Diffusion of gold (Au) over nickel solderable layer although initially dissolved away by soldering process. | Lack of solderability |
| Black pad formation [72] | Excessive amount of nickel corrosion (from gold-phosphorus reaction) during gold deposition process | |

Fig. 7 provides a sample qualification plan both for laser/SOA-PIC assembly, and assembly components. For assembly qualification plan, the laser/SOA-PIC samples dedicated for qualification are divided into three for mechanical stress, lifetime, and environment stress tests. Mechanical stress tests include wire-bond pull, wire-bond ball shear, and die shear tests [20], while environment stress tests include TMCL (~500 cycles), uBDH, HTS, and LTS [13, 19]. Lifetime tests are conducted by first preconditioning the laser/SOA-PIC samples via TMCL for ~20 cycles before performing optical measurements on the samples (e.g., retrieving the LIV curve, spectral and spatial profiles) before subjecting the samples to high-temperature high-bias burn-in test (e.g., 150°C, 500mA constantly applied for 48 hours) and re-performing the same optical measurements for anomaly detection to identify potentially defective devices. Samples that pass the test are either sent for product assembly packaging or sent for lifetests. Lifetest may be carried out by first dividing the samples into three groups subjected to different conditions (e.g., 100, 150, 200°C; and 300, 400, 500mA) to for statistical comparison in order to retrieve the aging factor [58]. For assembly component qualification plan, laser/SOA samples may be subjected to ESD tests while PIC samples — which has metallization layers, bond, wirebond, and wirebonding pads — may be subjected to back-end of line (BEOL) tests (e.g., SM and EM tests). To specifically test the PIC couplers, optical power handling and mechanical tests (vibration, mechanical shock, etc.) may be carried out.

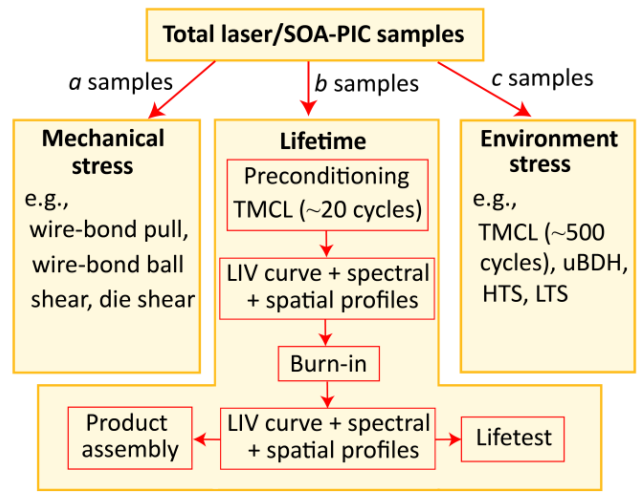

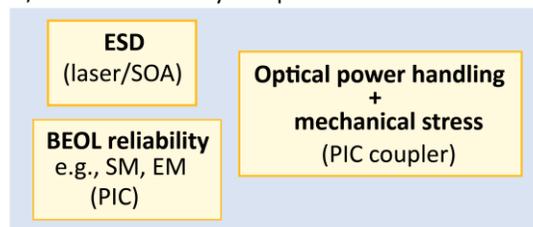

Fig. 7. Reliability qualification testing plan for (a) laser/SOA-PIC assembly, and (b) laser/SOA-PIC assembly components.

## VI. CONCLUSION

As there is currently no practical silicon-based optical source, the importance of laser/SOA-PIC integration to full-fledged implementation of silicon PIC cannot be understated. The complementary properties of silicon and laser/SOA materials (e.g., III-V materials) respectively suited for scaled-up optical circuitry and light generation have led to the development of various laser/SOA-PIC integration schemes. Despite its increasing maturity, there remains ample room for improvements in the integration technology. Research and development directions for such improvements may be established on the basis of identifying ideal/desirable integration properties. This includes high optical coupling efficiency, alignment tolerance, mechanical stability and optical power handling; efficient on-chip thermal management control, high aging and failure resistance. As the industry continues to dynamically grow in practical importance, the role of photonic foundries to develop and offer laser/SOA-PIC integration-related PDKs to the industry remains highly vital to the ecosystem of production, packaging, and design houses.

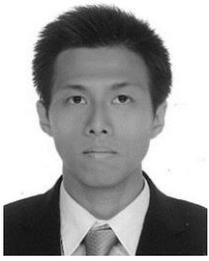

**James Y.S. Tan** received his B. Eng (Electrical) from University of Malaya in 2013, MSc. in Electrical Engineering from Korea Advanced Institute of Science and Technology (KAIST) in 2015, and DPhil. in Materials from University of Oxford in 2021. From 2015-2016, he was a researcher at KAIST, where he studied non-Hermitian physics, photon antibunching, and resonators. During his DPhil studies, he worked on optical neuromorphic computing. His current research interests include integrated photonic devices and systems.

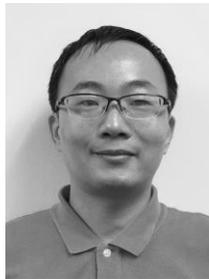

**Shawn Wu Xie** is Principal R&D Engineer in Advanced Micro Foundry Pte. Ltd. (AMF). He has seventeen years of industry experience in optical sub-assembly design, product characterization and manufacturing yield improvement. At AMF, his main research focus is on the study of silicon photonics application in optical transceiver module, including device characterization, photonics packaging and reliability validation. He received the B. Eng degree from Huazhong University of Science and Technology (HUST) and Master's degree in Optics from Shenzhen University in 2003 and 2006 respectively.

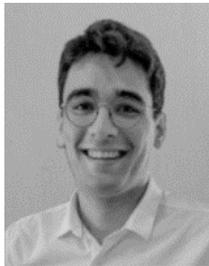

**Salih Yanikgonul** received the B.Sc degree from Bogazici University, Türkiye, in 2014, and the PhD degree in Electrical and Electronic Engineering from Nanyang Technological University (NTU), Singapore, in 2021. During the Ph.D. degree, he was engaged in research on single photon detection and manipulation in photonic integrated circuits with NTU and Institute of Materials Research and Engineering (IMRE), Agency for Science, Technology and Research (A*STAR), Singapore. Since 2020, he has been a Research Scientist with Advanced Micro Foundry Pte Ltd. (AMF), Singapore. His research interests include developing integrated photonic devices and circuits on silicon photonics and various integration platforms. He was the recipient of the Institute of Physics Singapore Outstanding Poster Award in 2019 and the Republic of Türkiye Prime Ministry Outstanding Achievement Scholarship.

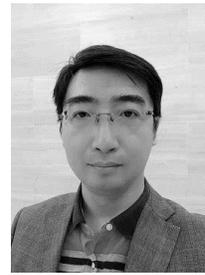

**Chao Li** received his B.S. and M.S. degrees in physics from Tsinghua University, Beijing, China, in 1998 and 2001, respectively. He received his Ph. D. degree in electrical and electronic engineering from Hong Kong University of Science and Technology (HKUST), Hong Kong in 2007. From 2007 to 2010, Dr. Li was a post-doctoral researcher in the Chinese University of Hong Kong (CUHK), where he was responsible for management and technical leadership in the developing of silicon-based passive and active optical devices for optical communication network. From 2010 to 2017, he was a Scientist with the Institute of Microelectronics, Agency for Science, Technology and Research (A*STAR), Singapore, where he was leading the development of silicon photonics packaging technology and platform to build low-loss, low cost, and low-profile silicon PIC modules. He is now a Co-Founder, Director at Advanced Micro Foundry Pte. Ltd. (AMF), focusing on silicon photonics' industrialization and commercialization. Dr. Li has contributed more than 80 journal articles, book chapters and conference talks. He has filed more than 20 IP/KHs.

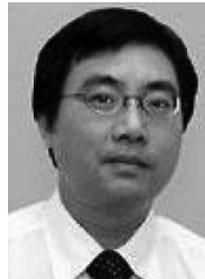

**Patrick Guo-Qiang Lo** received the M.S. and Ph.D. degrees in electrical and computer engineering from the University of Texas at Austin, USA, in 1989 and 1992, respectively. He was with Integrated Device Technology, Inc., both in San Jose, CA, and Hillsboro, Oregon, USA, from 1992 to 2004. He was involved in CMOS manufacturing areas in process and integration research and development. From 2004 to 2017, he was with the Institute of Microelectronics (IME), Singapore, where he was the Program Director of the Nanoelectronics and Photonics Program and TSV Taskforce. His current research interests include novel semiconductor device and integration technology, in the areas of nanoelectronics, Silicon microphotonics, GaN-based power electronics, and also emerging memory, particularly in the paths towards to productization and commercialization. Since 2017, he has been with Advanced Micro-Foundry Pte. Ltd. (AMF), Singapore focusing on silicon photonics' industrialization and commercialization. He has authored or coauthored more than 200 peer-reviewed journal and conferences publications, and holds more than 40 granted U.S. patents, he was a recipient of the IEEE George E. Smith Award in 2008 for the best paper published in the IEEE Electron Device Letters in 2007, and recipient of Singapore's National Technology Award in 2008 and President Technology Award in 2010.